\documentclass[aps,pre,amsmath,superscriptaddress,amssymb,reprint,floatfix]{revtex4-1}     
\usepackage{graphics}
\usepackage{color}
\usepackage{hyperref} 
\usepackage{epstopdf}
\usepackage{ifpdf} 
\usepackage{epsfig}
\usepackage{lipsum}

\begin{document}

\title{Confinement-induced alternating interactions between inclusions in an active fluid}

\author{Mahdi Zarif}
\email{m\_zarif@sbu.ac.ir}
\affiliation{Department of Physical and Computational Chemistry, Shahid Beheshti University, Tehran 19839-9411, Iran}

\author{Ali Naji}
\email{a.naji@ipm.ir}
\affiliation{School of Physics, Institute for Research in Fundamental Sciences (IPM), Tehran 19395-5531, Iran}


\begin{abstract}
In a system of colloidal inclusions suspended in a thermalized bath of smaller particles, the bath engenders an attractive force between the inclusions, arising mainly from entropic origins, known as the depletion force. In the case of active bath particles, the nature of the bath-mediated force changes dramatically from an attractive to a repulsive one, as the strength of particle activity is increased. We study such bath-mediated effective interactions between colloidal inclusions in a bath of self-propelled Brownian particles, being confined in a narrow planar channel. Confinement is found to have a strong effect on the interaction between colloidal particles, however, this mainly depends on the colloidal orientation inside the channel. Effect of the confinement on the interaction of colloidal disk is controlled by the layering of active particles on the surface boundaries. This can emerge as a competitive factor, involving the tendencies of the channel walls and the colloidal inclusions in accumulating the active particles in their own proximity. 
\end{abstract}


\maketitle

\section{Introduction}

Active matter is a nonequilibrium branch of soft matter and has risen strong interest in recent years. Self-propelled particles (or, swimmers) have their own engines to self-propel them in fluid media in the absence of external forces in a directed motion \cite{purcell:AJP1977,berg:book2003,Lauga:RPP009,ramaswamy:ARCMP2010,golestanian:SoftMatter2011,lauga:SoftMatter2011,Romanczuk:EPJ2012,Marchetti:RMP2013,Yeomans:EPJ2014,Elgeti:RPP2015,cates:ARCMP2015,Goldstein:ARFM2015,Lauga:ARFM2016,Zottl:JPCM2016,bechinger:RMP2016}. However, in the presence of torque, the line of the motion is no longer aligned with that of the self-phoretic force and the swimmer tends to execute a circular motion (known as chiral swimmers) \cite{Brokaw:JEB1958,Brokaw:JCCP1959,Teeffelen:PRE2008,Friedrich:PRL2009,Kummel:PRL2013,Volpe:AJM2013,Crespi:PRE2013,Volpe:AJP2014,Boymelgreen:PRE2014,Reichhardt:PRE2014,Breier:PRE2014,Li:PRE2014,Mijalkov:SoftMatter2015,Xue:EPL2015,Crespi:PRL2015,Lowen:EPJST2016,Zaeifi:JCP2017,Ai:SoftMatter2018}. In general, these systems often exhibit athermal (or active) Brownian motion and the self-propulsion force dominates random thermal fluctuations. In nature biological machines are abundant and they are able to transport chemical energy into a directed motion using their own motor \cite{berg:book2003,Goldstein:ARFM2015,SHACK:BMB1974,Woolley:Reproduction2003,SELLERS:BBA20003,Roberts:NRMCB2013,Hirokawa:NRMCB2009}. Examples of self-propelled particles are biological motors \cite{SELLERS:BBA20003,Roberts:NRMCB2013,Hirokawa:NRMCB2009}, 
multiflagellate bacterium {\em E. coli} \cite{berg:book2003}, biflagellate alga {\em C. reinhardtii} \cite{Goldstein:ARFM2015} and uniflagellated sperm cells  \cite{SHACK:BMB1974,Woolley:Reproduction2003}. Also, recent advances in experimental techniques have enabled fabrication of artificial nano-/swimmers (e.g., active Janus particles) \cite{Perro:JMC2005,Mano:JACS2005,Dhar:NL2006,Walther:SoftMatt2008,Jiang:AM2010,Douglass:NP2012,Walther:CR2013,Buttinoni:PRL2013,Volpe:SR2014,Bianchi:SR2016,Poggi:CPS2017,Zhang:langmuir2017}.

The performance of swimmers is often affected by the presence of boundaries, interfaces and other particles. It has been proven that the swimmers (biological, synthetic or simulated) tends to swim and accumulate close to the boundaries \cite{ROTHSCHILD:Nat1963,Biondi:AIChE2004,Mannik:PNAS2009,Lebleu:JMS2009,binz:ME2010,Wensink:EPJST2013,Ao:EPJST2014,Soto:PRE2014,Elgeti:RPP2015,Malgaretti:JCP2017}. Nature has many examples of moving biological swimmers in confined regions e.g. sperm cells in female reproductive tract \cite{Suarez:hpd2006} and technological applications of microfluidic devices \cite{Barbot:SR2016}. These situations have been studied in a more systematic way in recent years by theoretical \cite{Lowen:JPCM2001,Kumar:EPL2012,Lee:NJP2013,Costanzo:EPL2014,Malgaretti:JCP2017} and experimental \cite{Yu:Small2016,Wu:JACS2017,Yang:PNAS2017} techniques. When a self-propelled particle collides with an interface, this is a purely steric interaction and as a result, its normal component of the self-propulsion velocity is canceled due to hard-core repulsive force. This causes to particle slides along the interface until its rotational diffusion coefficient changes the orientation of the particle. However, rotational coefficient still can reorient the particle toward the interface which can lead to particle stays close to interface even for a longer period of time.

By putting {\em nonactive} macromolecular (e.g., colloidal) particles (inclusions) in a medium filled by smaller (e.g., polymers) particles (depletants), causes an induced effective, short ranged, attractive interaction between them \cite{Lekkerkerker:book2011,Likos:PR2001,Likos:PRL2003,Angelani:PRL2011,Harder:JCP2014,Cacciuto:PRE2016,Cacciuto:SoftMatter2016,Leite:PRE2016,Zaeifi:JCP2017,Dolai:SoftMatter2018,Yang:JPCM2018}. This force is mainly dependent on the shape and propulsion of the bath particles. For example, in the case of colloidal disk immersed in a bath of disk-shaped nonactive particles (no self-propulsion) this force is a purely attractive force that has an entropic origin. However, by introducing particle self-propulsion, this effective force changes to an induced short-range repulsive interaction \cite{Harder:JCP2014,Zaeifi:JCP2017}. In this case, the surface of the large inclusions serves as a spatial confinement for the suspending smaller particles. In our previous work \cite{Zaeifi:JCP2017} we show that the chirality of the {\em active} particles changes this short-ranged attractive force into an induced repulsive force. Here, the formation of circular layers (or \emph{`rings'}) of active particles around the colloidal inclusions plays a critical role and the effective force induced is mainly determined by the interactions of these rings.

Depletion interactions have been studied extensively, both experimentally and theoretically,  in the context of nonactive macromolecular mixtures in equilibrium that is restricted to a confined geometry \cite{Asakura:JCP1954,Trokhymchuk:Langmuir2001,Xiao:EPL2006,Chervanyov:PRE2011,Rahul:SoftMatter2013,Usatenko:EPJST2017,Lekkerkerker:book2011}. However, these interactions are yet to be known for the case of active particles. In this paper, we tackle this problem using a simplified and standard model of disk-shaped inclusions in an active bath within a commonly used, two dimensional model of self-propelled particles constrained to a narrow channel. By introducing the wall into a mixture of colloidal inclusions in an active bath, apart from the formation of the circular layers around the colloids, also, due to the interactions between surface and the swimmers, layers of swimmers exist close to the wall. The sequential overlaps between the colloidal rings and the wall layers generate the distinct features of the force profiles. In a competition between colloids and the channel wall as a site for active particles gathering places, because of the convex curvature of the colloidal surfaces, the layers of particles are more populated for the wall. However, the density of layers has a strong dependence on the swim speed. We investigate these parameters along with the orientation of colloidal inclusions on effective  interactions mediated between colloidal disks.

In our previous work \cite{Zaeifi:JCP2017}, we investigated  the effect of layer formation on the interactions between colloidal inclusions in a bath of active particles and focused on the regime of low occupied area fractions. Here, we take a similar direction, however, we confine the system into a narrow channel to study the effects of the channel confinement and consider both cases of relatively low and relatively high area fractions but, in any case, below the onset of motility-induced phase separation \cite{Gompper:SM2018}. Since the channel introduces an anisotropy on the system, we consider two different orientation for the colloidal inclusions, with their center-to-center axis being parallel or perpendicular to the channel walls.

The rest of this paper is organized as follows: In Section~\ref{sec:Model}, we introduce our model and details of the numerical simulation techniques used in this work by focusing on a standard two-dimensional model of self-propelled particles. In Section~\ref{sec:results}, we discuss the results for the depletion interaction between two disk-shaped inclusions in a bath of self-propelled swimmers at two different colloidal orientations, Section~\ref{sec:Vert} for the parallel alignment and Section~\ref{sec:Horz} for the perpendicular alignment. Both cases are studied at two different area fractions of bath particles. The paper is concluded in  Section~\ref{sec:Conclusion}.

\section{Model and methods}
\label{sec:Model}

Our two-dimensional model consists of a pair of identical, impenetrable and nonactive colloidal inclusions of diameter $a_c$ confined within a narrow channel of height $H$, containing $N$ identical, self-propelled Brownian particles (swimmers) of smaller radius $a$ at area fraction $\phi$. 

Active particles are self-propelled at constant speed $V_s$ along a direction specified by an angular coordinate, $\theta$, hence, their swim orientation is characterized by the unit vector ${\mathbf n}=(\cos \theta, \sin \theta)$ in within the $x-y$ plane. The particles are subjected to both translational and rotational diffusion with coefficients $D_T$ and $D_R$, respectively. We ignore active noise effects and assume that the diffusive processes are thermal in nature; therefore, $D_T$ is related to particle (translational) mobility through the Einstein relation, $D_T  = \mu_T k_{\mathrm{B}}T$, where $k_{\mathrm{B}}$ is the Boltzmann constant and $T$ is the ambient temperature. Also, for no-slip sphere in the low-Reynolds-number regime, as supposed for the model particles here, the translational and rotational diffusion coefficients are related as $D_R=3D_T/4a^2$ \cite{happel:book1983}. The dynamics of the active particle position, ${{\mathbf r}}_i(t)$, and orientation, ${\theta}_i(t)$,  obey the overdamped Langevin equations
\begin{eqnarray}
\dot{{\mathbf r}}_i &= &V_s{\mathbf n}_i-\mu_T\frac{\partial  U(\{{\mathbf r}_j\})}{\partial {{\mathbf r}_i}}+\sqrt{2 D_T}\, {\boldsymbol \eta}_i(t),
\label{Eq:langevin_a}
\\
\dot{\theta}_i&=& \sqrt{2D_R}\, \zeta_i(t),
\label{Eq:langevin_b}
\end{eqnarray}
where ${\boldsymbol \eta}_i(t)$ and $\zeta_i(t)$ are independent Gaussian white-noise terms, representing random force and random torque respectively, with zero mean,  $ \langle { \eta}_i^\alpha(t)  \rangle=  \langle \zeta_i(t)  \rangle=0$, and two-point time correlation functions $ \langle { \eta}_i^\alpha(t) { \eta}_j^\beta(t') \rangle=\delta_{ij}\delta_{\alpha\beta}\delta(t-t')$ and  $ \langle \zeta_i(t) \zeta_j(t') \rangle=\delta(t-t')$, with $i, j=1,\ldots,N$ and $\alpha, \beta=1, 2$, denoting the Cartesian coordinates. $U$ in Eq. (\ref{Eq:langevin_a}) is the sum of pair potentials acting between any pairs of particles in the system. The particle-particle $(V({\mathbf r}_{ij}))$ and particle-wall interactions $(V_{\mathrm{W}}^\pm(y_i)$ with $\pm$ denoting the top and bottom walls)  are modeled using the purely repulsive pair potentials according to a conventionally modified form of the Weeks-Chandler-Andersen (WCA)
potential, i.e., 
\begin{equation}
 V({\mathbf r}_{ij}) =
\left\{\begin{array}{ll}
4\epsilon\! \left [ \left ( \frac{\sigma_{\mathrm{eff}}}{|{\mathbf r}_{ij}|} \right )^{12} -2 \left ( \frac{\sigma_{\mathrm{eff}}}{|{\mathbf r}_{ij}|} \right )^{6}+1\right ] &\,\, |{\mathbf r}_{ij}| \leq \sigma_{\mathrm{eff}}, \\ 
\\
0 &\,\, |{\mathbf r}_{ij}| > \sigma_{\mathrm{eff}},
\end{array}\right.
\end{equation}
and 
\begin{equation}
 V_{\mathrm{W}}^\pm(y_i) =
\left\{\begin{array}{ll}
4\epsilon\! \left [ \left ( \frac{a}{|\delta y_i^\pm|} \right )^{12} -2 \left ( \frac{a}{|\delta y_i^\pm|} \right )^{6}+1\right ] &\,\, |\delta y_i^\pm| \leq a, \\ 
\\
0 &\,\, |\delta y_i^\pm| > a,
\end{array}\right.
\end{equation}
respectively. Here, $|{\mathbf r}_{ij}|$ is the center-to-center distance between any two particles with $\sigma_{\mathrm{eff}}=2a$, when the pair of particles considered are both of active particles, $\sigma_{\mathrm{eff}}=2a_c$, when the pair are both of colloidal inclusions, and $\sigma_{\mathrm{eff}}=a+a_c$, when one of the particles considered is an active particle and the other one is a colloidal inclusion. Also, $\delta y_i^\pm=y_i\mp H/2$ is the  perpendicular distance of the particle from the top/bottom walls. In all  cases, it is assumed that the interaction energy strength is given by a single parameter, $\epsilon$. 

\begin{figure}
 \begin{center}
 \includegraphics[width=3.0in]{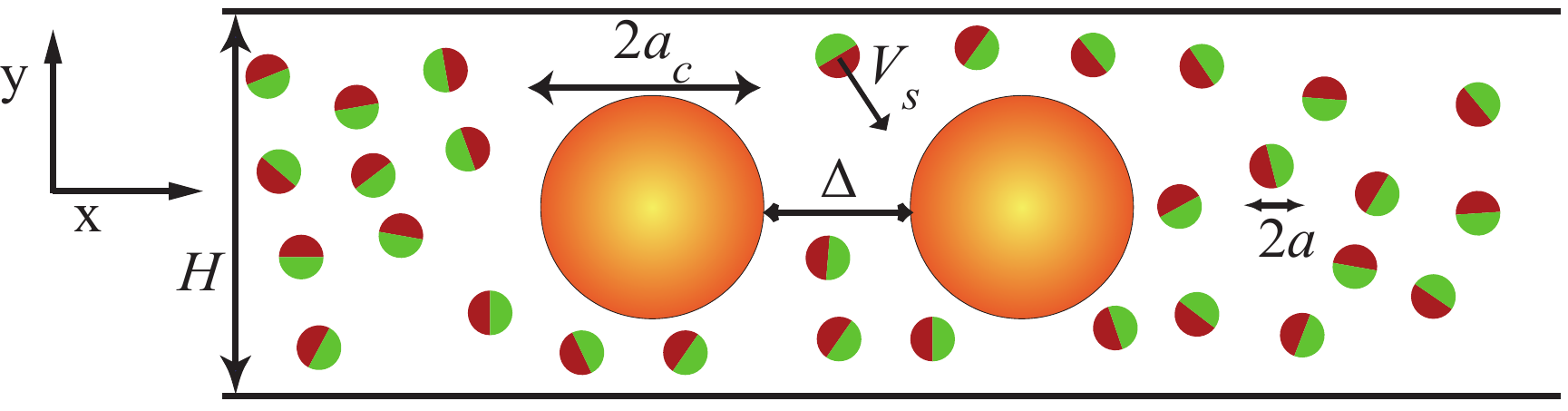}
 \caption{Schematic view of two fixed, identical, impenetrable, and nonactive colloidal disks of radius $a_c$  at surface-to-surface distance $\Delta$ inside a channel of height $H$, containing active particles of radius $a$ and self-propulsive speed $V_s$. 
 }
 \label{fig:schem_colloids}
\end{center}
\end{figure}

Our assumption of thermal noises through the Langevin equations ensures that  the system approaches its appropriate Boltzmann-weighted equilibrium determined by the potential $U(\{{\mathbf r}_j\})$ in the stationary state, when the active forces are switched off. 

We use a dimensionless representation in our simulations by rescaling the units of length and time with the radius of the active particles and the corresponding time-scale for translational diffusion as
\begin{equation}
\tilde x = \frac{x}{a},\quad \tilde y = \frac{y}{a},\quad \tilde t = \frac{D_T t}{a^{2}},\quad \tilde H = \frac{H}{a}.
\label{Eq:dimensionless}
\end{equation}

Equations (\ref{Eq:langevin_a}) and (\ref{Eq:langevin_b}) can be solved numerically using Brownian Dynamics simulations by rewriting them in dimensionless and discrete forms for time-evolution over a sufficiently small time-step, $\Delta \tilde t$, as 
\begin{eqnarray}
&&\tilde x_i(\tilde t +\Delta \tilde t ) = \tilde x_i(\tilde t)+[Pe_s \cos \theta_i(\tilde t) + \tilde f^x_i(\tilde t)] \Delta \tilde t + \sqrt{2 \Delta \tilde t}\,R^x_i\nonumber\\
\\
&&\tilde y_i(\tilde t +\Delta \tilde t ) = \tilde y_i(\tilde t)+[Pe_s \sin \theta_i(\tilde t) + \tilde f^y_i(\tilde t)] \Delta \tilde t  + \sqrt{2 \Delta \tilde t}\,R^y_i\nonumber\\
\\
&&\theta_i (\tilde t +\Delta \tilde t ) = \theta_i(\tilde t) + \sqrt{2 \chi \Delta \tilde t}\,R^\theta_i,
\label{Eq:2Ddimensionless}
\end{eqnarray}
where $\tilde f^x_i = -\partial \tilde U/\partial \tilde x_i$ and  $\tilde f^y_i = -\partial \tilde U/\partial \tilde y_i$ are the Cartesian components of the dimensionless force derived from the rescaled potential $\tilde U=U/(k_{\mathrm{B}}T)$, and $R^x_i$,  $R^y_i$ and $R^\theta_i$ are independent, Gaussian random numbers with zero mean and unit variance.  We have defined  $\chi = a^2D_R/D_T = 3/4$, which,  as noted before, follows from the standard relations for diffusion coefficients of spherical particles \cite{happel:book1983}. 
We also define the rescaled swim P\'eclet number,
\begin{equation}
Pe_s=\frac{a V_s}{D_T}, 
\end{equation}
as the ratio of the characteristic time-scale of translation diffusion, $a^2/D_T$, and that of the particle swim, $a/V_s$. 

In the rescaled representation, the system is described by the size ratio, $a_c/a$, the rescaled area fraction, $\phi a^2$, the swim P\'eclet number, $Pe_s$, and the rescaled center-to-center distance between adjacent colloidal inclusions, $d/a$, or equivalently, their rescaled surface-to-surface distance, $\tilde \Delta = \Delta/a$ (see Fig. \ref{fig:schem_colloids}). In what follows, we use the fixed parameter values $\epsilon/(k_{\textup{B}}T)=10$, $\phi a^2 = 0.1$ and $0.3$, $a_c/a = 5$  and consider only two and three fixed colloidal inclusions in the simulated active bath (the more general cases with many fixed or mobile colloidal inclusions will be considered elsewhere \cite{Zaeifi:JCP2017}). The swim P\'eclet number is increased from $Pe_s=0$ up to  $Pe_s=50$. Typical rescaled parameter values such as $Pe_s = 5$ can thus be mapped to a wide range of experimentally accessible, actual parameter values such as $a = 1\, \mu{\mathrm{m}}$, $V_s = 10\, \mu{\mathrm{m}}\cdot{\mathrm{s}}^{-1}$, $D_T \simeq 0.22\, \mu{\mathrm{m}}^2\cdot{\mathrm{s}}^{-1}$, $D_R \simeq 0.16\,{\mathrm{s}}^{-1}$, and $\eta = 0.001\, {\mathrm{Pa\cdot s}}$.

Our simulations are performed using $N = 400$ active bath particles distributed initially in random positions in the channel bounded laterally (along the $x$-axis) by a box of lateral size $\tilde {L}= 4 N \pi / \phi \tilde H$, where we impose periodic boundary conditions.The simulations typically run for $10^6-10^8$ time steps (with an initial $10^6$ steps used for relaxation purposes) and averaged quantities are calculated over a sufficiently large statistical sample (by choosing different initial conditions) after the system reaches a steady state.

One of the key quantities that we shall use later in the text is the net force acting on colloidal inclusions applied by individual bath particles while colliding with them. This force is calculated using averaged sum of forces ($\tilde f^x_i$ and $\tilde f^y_i$) applied by swimmers-colloid collisions. The bath of our system is homogeneous and isotropic, therefore, for a single colloidal inclusion, the {\em net} force due to the bath particle collisions will be averaged to zero (the force applied from either side of colloid will be equal in magnitude and cancel out each other). However, in the case of two colloidal inclusions, depending on their orientation, the isotropicity of the bath along the center-to-center axes of colloids breaks and as a results a {\em net} force is exerted on colloidal inclusions. This non-zero force is interpreted as effective two-body force and is defined as $\tilde {\mathbf F}_2  =  \tilde F_2 \hat{\mathbf x}$. This quantity is calculated in our simulations using averaging over different configurations using $\tilde {\mathbf F}_2 = \sum_i \langle\tilde {\mathbf f}_i\rangle$, where the brackets $\langle\cdots\rangle$ represents the statistical average over different configurations and $\tilde {\mathbf f_i}=(\tilde f_i^x, \tilde f_i^y)$ is the force applied to the $i$th particle by neighbor particles that collides with it via the WCA interaction potential. For the sake of demonstration, we have divided the force $\tilde {\mathbf f_i}$ by swim P\'eclet number, $Pe_s$, which to include the case where $Pe_s=0$ we have conventionally chosen to divide the effective force values by $Pe_s+1$. This defines a new quantity defined as $\hat F_2 \equiv  \tilde F_2/(Pe_s+1)$.

\section{Results}
\label{sec:results}

Here, we will focus on two different orientational configurations for the fixed colloidal inclusions inside the channel, i.e., with their center-to-center line being either parallel or perpendicular to the $x$-axis, in such a way that the up-down symmetry with respect to the centerline of the channel is preserved. In each case, we select two different area fractions, $\phi a^{2} = 0.1$ and 0.3, corresponding to low and high  area fractions (and, in any case,  below the onset of motility-induced phase separation \cite{Gompper:SM2018}). 
\\

\subsection{Parallel orientation}
\label{sec:Horz}

\subsubsection{Low area fraction $(\phi a^{2} = 0.1)$}
\label{sec:rho0p1Horz}

\begin{figure}
\centering
\includegraphics[width=1.0\linewidth]{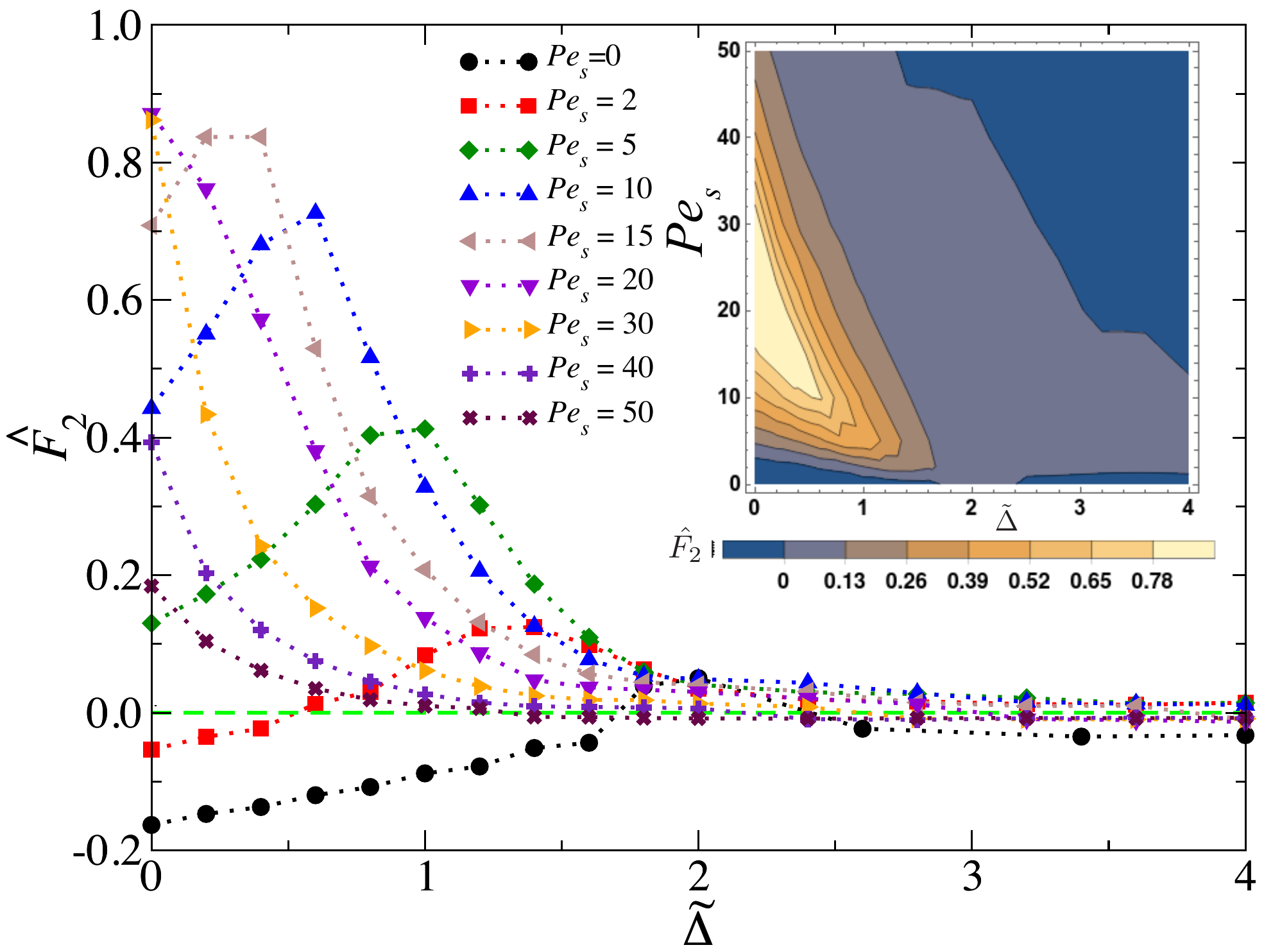}
\caption{Rescaled effective two-body force, $\hat F_2$, acting on each of the two colloidal inclusions immersed parallelly in an active bath with area fraction $\phi a^{2} = 0.1$ is shown as a function of their rescaled surface-to-surface distance, $\tilde \Delta$, at fixed rescaled channel height of $\tilde H = 15$ and for different values of the P\'eclet number, $Pe_s$, as indicated on the graph. Note that the plot shows effective force divided by $Pe_s+1$ as defined in the text. Symbols are simulation data and lines are guides to the eye. Inset: Contour plot of $\hat F_2$ as a function of the system parameters $Pe_s$ and $\tilde \Delta$. The magnitude of attractive and repulsive forces are separated with different contour lines and colors.
}
\label{fig:fig2}
\end{figure}

\begin{figure}[t!]
\centering
\includegraphics[width=1.0\linewidth]{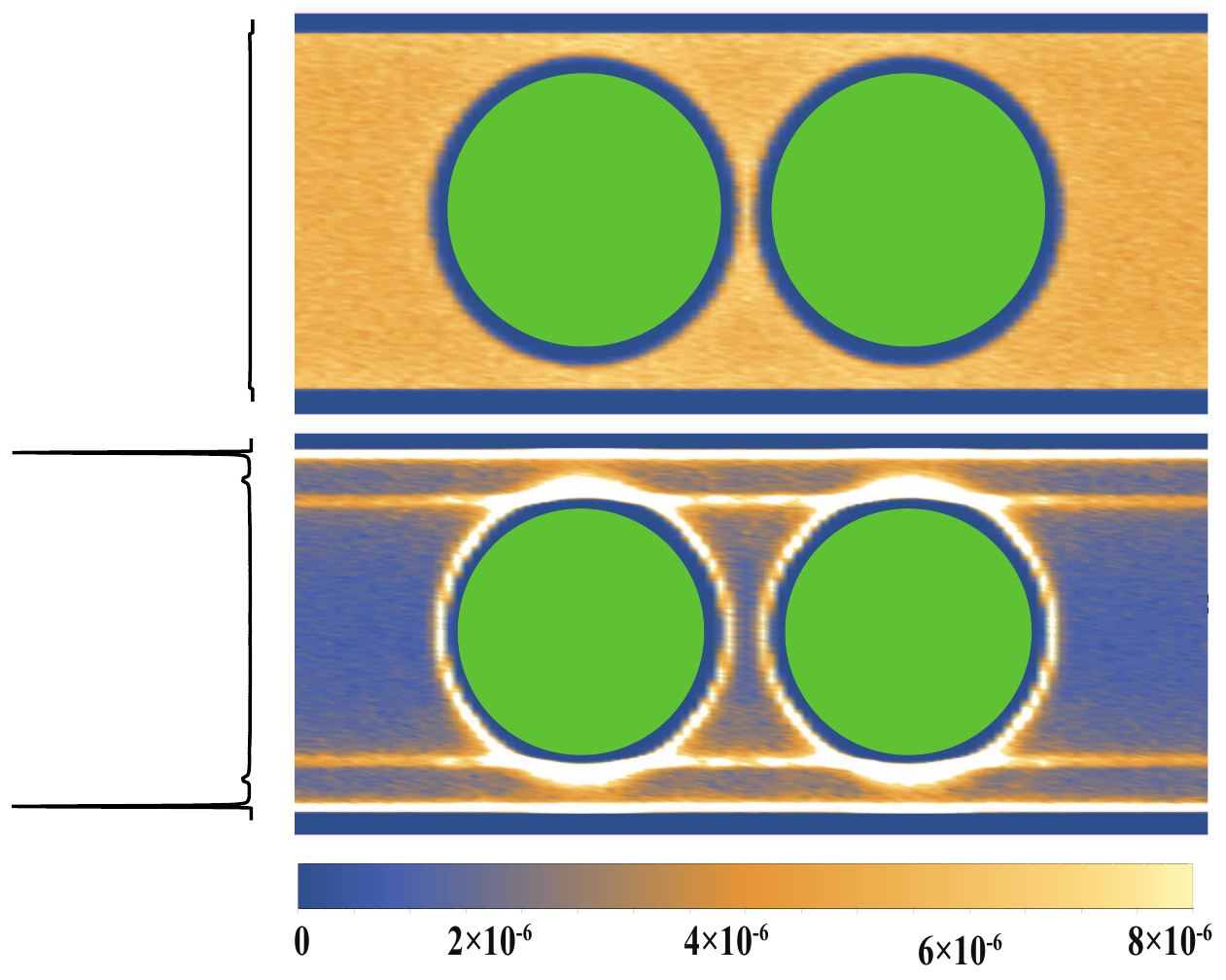}
\caption{Steady-state particle density maps for $Pe_s = 0$ (top) and $Pe_s = 50$ (bottom) around two colloidal inclusions located parallelly with $\tilde \Delta = 2$, $\tilde H = 15$  and $\phi a^{2} = 0.1$.
} 
\label{fig:fig3}
\end{figure}

Figure \ref{fig:fig2} shows our simulation data for the rescaled effective two-body force acting on each of the colloidal inclusions as a function of the colloidal surface-to-surface distance $(\tilde \Delta)$ at different P\'eclet numbers and for a fixed rescaled channel height of $\tilde H = 15$. Because of the symmetric positional arrangement of the inclusions within the channel, the net (statistically averaged) force acting on the two disks are equal with the $x$ (parallel-to-channel) component averaging out to be zero. Thus, Fig.~\ref{fig:fig2} shows only the $y$ component of the rescaled net force acting on each of the inclusions due to the active particles. At vanishing and small swim P\'eclet numbers $Pe_s$, similar to the passive case, the rescaled force is attractive \cite{Lekkerkerker:book2011}, over short distances where the depletion layers  (thin, dark-blue, circular regions in Fig. \ref{fig:fig3}, top)   around the colloids can overlap.  However, by increasing the P\'eclet number of the bath particles, the rescaled force exerted on the colloidal disks changes to a repulsive one. The repulsive force profiles exhibit nonmonotonic behaviors as functions of the separation between the colloids, but with distinctly different features than those found in the nonconfined (bulk) geometry, where the two disks are immersed in an unbounded active bath \cite{Harder:JCP2014,Zaeifi:JCP2017}. As it was shown in the bulk case \cite{Zaeifi:JCP2017}, the nonmonotonicity of the interaction profiles is linked with  swimmer structuring around the colloidal disks; i.e., the swimmers form ring-like, high-density regions around the colloidal disks and the sequential overlaps between these rings from one colloid with the surface of the other, as the distance between the disks is decreased, resulting in nonmonotonic changes (alternating or oscillating rise-and-fall behavior) in the force profiles.

In a confined channel (Fig.  \ref{fig:fig2}), we find the following qualitative differences relative to the bulk system at the same area fraction: First, rather than the typical alternating rise-and-fall behaviors (see Fig. 2 in Ref. \cite{Zaeifi:JCP2017}, and Section \ref{sec:rho0p3Horz} below), here we find a single, relatively broad peak at short surface-to-surface separations. Second, the typical force magnitude increases by P\'eclet number until it reaches a maximum and decreases afterward; this occurs as the location of the peak in the force profiles moves to smaller separations. These two features are  more clearly discerned through the color-coded contour plots of the force profiles in the $Pe_s-\tilde \Delta$ plot shown in the inset of Fig. \ref{fig:fig2}. They reflect the fact that, unlike the bulk system, most of the swimmers in the present case are absorbed by the bounding channel walls, because the typical swimmer `detention' times on a flat surface are larger than those on a convex surface \cite{Lowen:JPCM2001,Malgaretti:JCP2017}. Thus, at a given $Pe_s$, a smaller fraction of swimmers are bound to the colloids as compared with the bulk system, and only a single `ring' or swimmer layer forms around each of the colloids at the area fraction considered here (thin, white, circular regions in Fig. \ref{fig:fig3}, bottom); hence, a single hump develops in the force profiles due to the ring-colloid surface overlaps at small surface-to-surface separations \cite{Zaeifi:JCP2017}.

\begin{figure}[t!]
\centering
\includegraphics[width=1.0\linewidth]{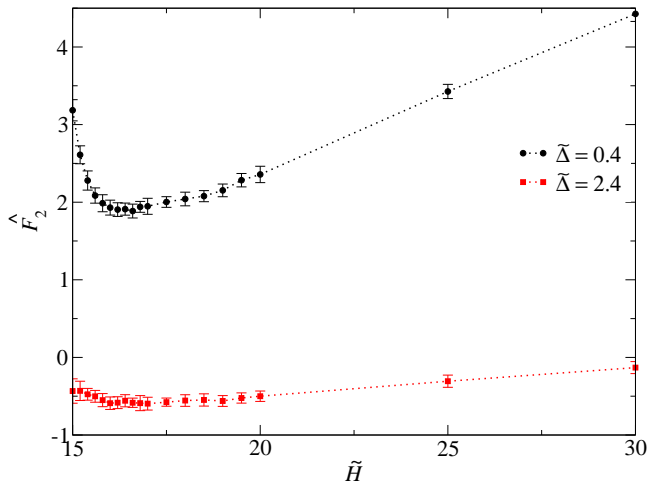}
\caption{Effective two-body force, $\hat F_2$, acting on each of the two colloidal inclusions immersed parallelly in an active bath is shown as a function of rescaled channel height, $\tilde H$, for  $Pe_s = 50$, $\phi a^{2} = 0.1$, and two fixed values of $\tilde \Delta$, as indicated on the graph. Symbols are simulation data and lines are guides to the eye.} 
\label{fig:fig4}
\end{figure}

Also, as the swim P\'eclet number is increased, the fraction of swimmers attracted to the channel wall increases making the proximity of the colloids less swimmer-populated and, as a consequence, the repulsive force is smaller in magnitude. As seen in Fig. \ref{fig:fig3}, the swimmer absorption by the channel walls leads to formation of more than one flat layers of swimmers next to the walls (here two layers shown by the white strips at each walls). These layers can even overlap with the rings formed around the colloids. To illustrate the effects of these boundary layers, we fix the position of the colloidal inclusions at two different surface-to-surface distances, $\Delta = 0.4$ and $\Delta = 2.4$, and plot the rescaled force,$\hat F_2$, acting on each of the colloids as a function of the channel width, $\tilde H$, in Figure~\ref{fig:fig4}. The force at  shorter colloidal distances is larger but the qualitative behavior of the force as a function of $\tilde H$ remains qualitatively the same and of a nonmonotonic form with a local minimum at around $\tilde H\simeq 16.2 $. This is the distance where the layers associated with walls avoid overlapping the rings associated with the colloids. For channel widths above the mentioned value, the force increases linearly with $\tilde H$. This is due to the fact that when the value of $\tilde H$ increases the length of the channel decreases. Due to the low density of swimmers only two layers exist around the walls and as a result, the density of swimmers in the bulk of the channel increases and this cause an enhancement of the rings around the colloids. At very larger channel widths, the force is expected to drop eventually and tend to its bulk value \cite{Harder:JCP2014,Zaeifi:JCP2017}, albeit to one with a smaller corresponding effective area fraction, as a fraction of swimmers remain indefinitely bound to the channel walls.

\begin{figure}
\centering
\includegraphics[width=1.0\linewidth]{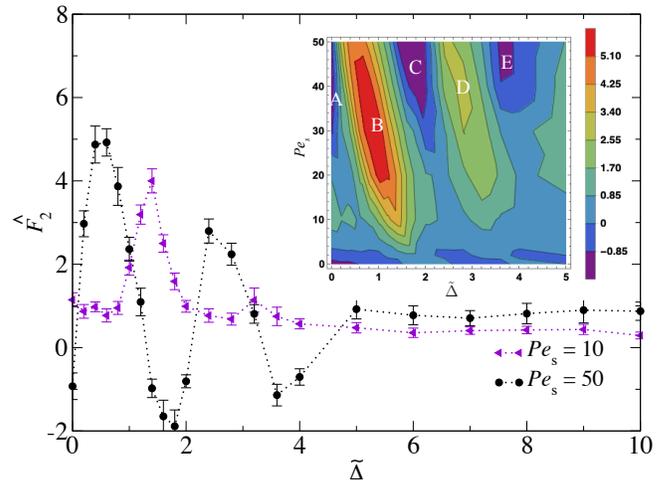}
\caption{Rescaled effective two-body force, $\hat F_2$, acting on each of the two colloidal inclusions immersed parallelly in an active bath with occupied area fraction of $\phi a^{2} = 0.3$ is shown as a function of  their rescaled surface-to-surface distance, $\tilde \Delta$, at fixed rescaled channel height of $\tilde H = 15$ for the cases $Pe_s=10$ and 50. Note that the plot shows effective force divided by $Pe_s+1$ as defined in the text. Symbols are simulation data and lines are guides to the eye. Inset: Contour plot of $\hat F_2$ as a function of the system parameters  $Pe_s$ and $\tilde \Delta$. The magnitude of attractive and repulsive forces are separated with different contour lines and colors.
}
\label{fig:fig5}
\end{figure}

\subsubsection{High area fraction $(\phi a^{2} = 0.3)$}
\label{sec:rho0p3Horz}

The area fraction occupied by the swimmers is another key parameter that significantly affects the two-body force mediated between two colloidal inclusion in a bath of active particles.  Figure \ref{fig:fig5} shows that the force behavior as a function of the colloidal surface-to-surface distance changes drastically at the higher area fraction of $\phi a^{2} = 0.3$, when compared to the behavior of the low-area-fraction system in Fig. \ref{fig:fig2} ($\phi a^{2} = 0.1$) with identical parameter values, i.e., $\tilde H=15$ and $Pe_s=50$. As seen, the force profile exhibit the alternating or oscillating behavior, which, as we noted in the preceding section, emerges at even lower area fractions in bulk systems \cite{Zaeifi:JCP2017}. Such a behavior is reflective of the formation of ringlike structures (circular layers with high swimmer density) around the colloids. Unlike the low-area-fraction case (Fig. \ref{fig:fig3}, bottom), here we find not only a primary but also a secondary ring. 
This is because a larger fraction of swimmers are available to be attracted to the colloidal surfaces at higher volume fractions. 

As the surface-to-surface distance between the colloids decreases, first the outer (secondary) rings of one colloid intersect with the surface of the other colloid and then the inner (primary) rings of one colloid intersect with the other colloid surface. These surface intersections give rise to the two peaks in the force profile in Fig. \ref{fig:fig5}; see Ref. \cite{Zaeifi:JCP2017} for further details.

\begin{figure}
\centering
\includegraphics[width=1.0\linewidth]{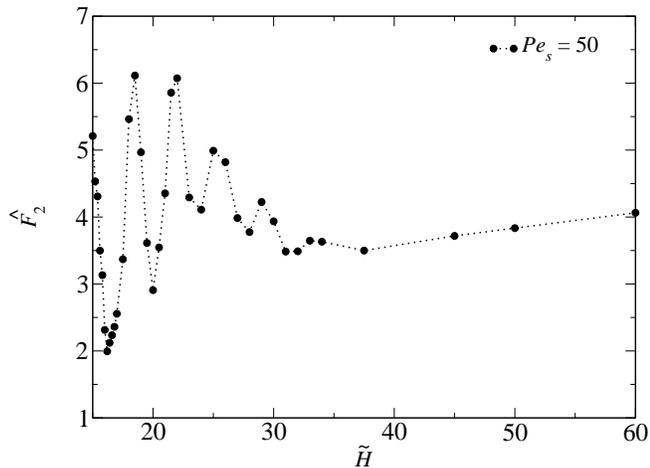}
\caption{Effective two-body force, $\hat F_2$, acting on each of the two colloidal inclusions immersed parallelly in an active bath with occupied area fraction of $\phi a^{2} = 0.3$ is shown as a function of rescaled channel height, $\tilde H$, for fixed $Pe_s = 50$ and $\tilde \Delta = 0.6$. Symbols are simulation data and lines are guides to the eye.
}
\label{fig:fig6}
\end{figure}

\begin{figure}
\centering
	\includegraphics[width=7cm]{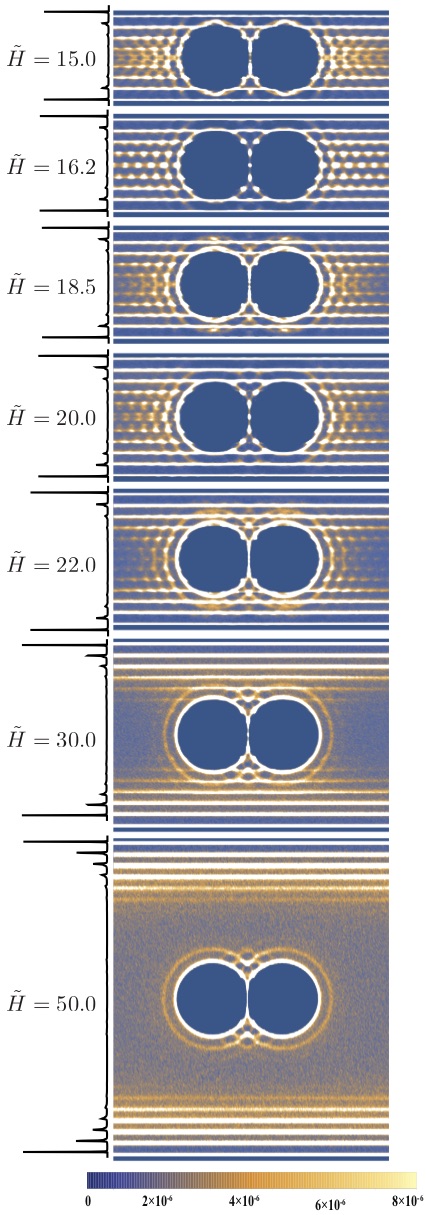}
	\caption{Steady-state density map of active microswimmers with $Pe_s = 50$, around two colloidal inclusions located parallelly with $\tilde \Delta = 2$ in a channel with different channel height, $\tilde H$, as indicated on the graph. The occupied area fraction of the active bath is $\phi a^{2} = 0.3$.}
	\label{fig:fig7}
\end{figure}

\begin{figure}
\centering
\includegraphics[width=1.0\linewidth]{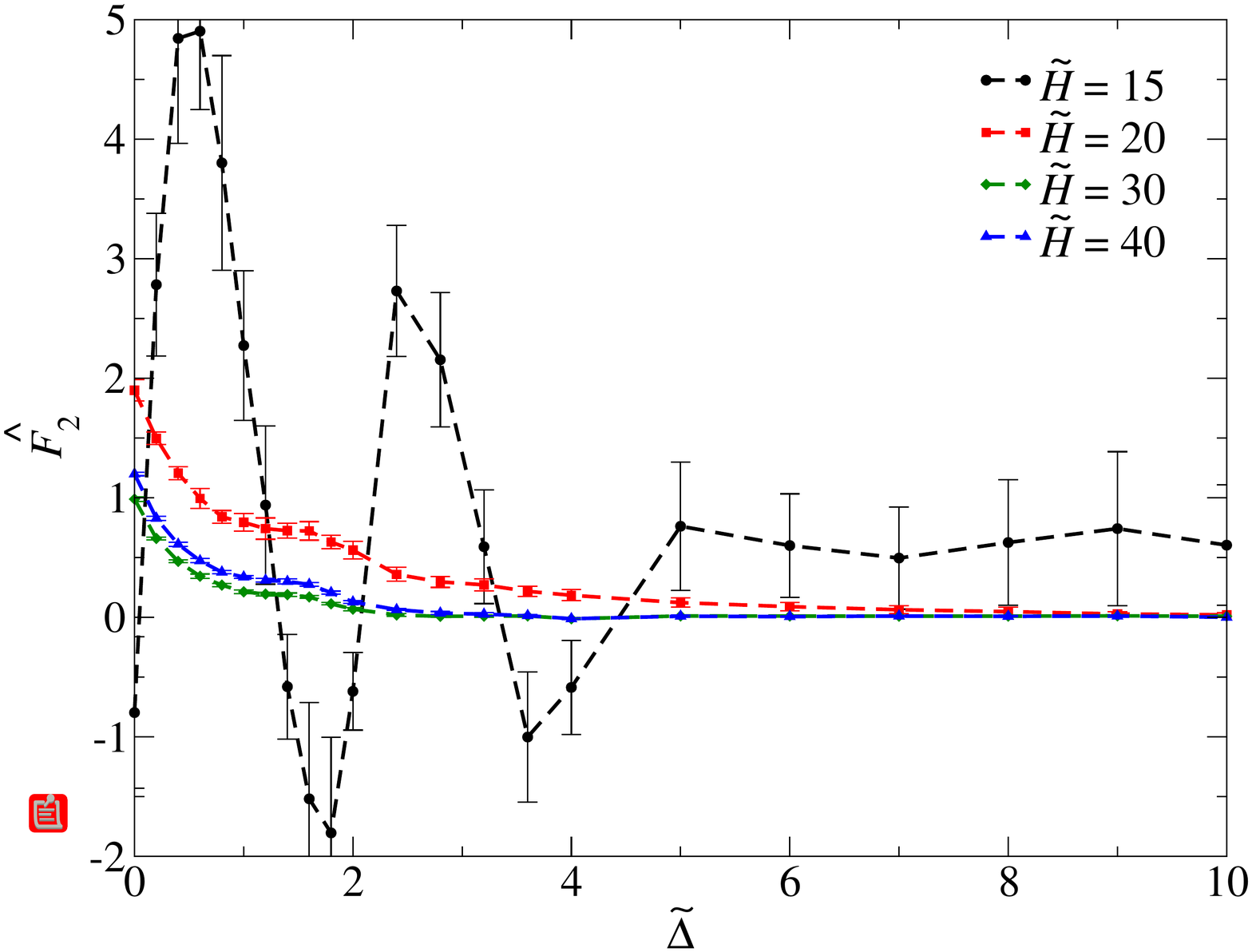}
\caption{Effective two-body force, $F_2$, acting on each of the two colloidal inclusions immersed parallelly in an active bath with fixed $\phi a^{2} = 0.3$, $Pe_s = 50$ and rescaled channel width, $\tilde H$,  values shown on the graph. Symbols are simulation data and lines are guides to the eye.
}
\label{fig:fig8}
\end{figure}

The alternating rise-and-fall behavior of the two-body force {$\hat F_2$ is shown across a wide range of $Pe_s$ in the force contour plot of Fig. \ref{fig:fig5}, clearly indicating that the said behavior starts to appear at relatively small P\'eclet numbers $Pe_s<10$ (note that the error bars in the plotted values of $\hat F_2$ is $\lesssim 0.5$). The positive (repulsive) peaks  (yellow/red regions, e.g., marked by B and D) are well developed already for $Pe_s>10$. The force falls down to negative (attractive) minimum values (purple regions, e.g., marked by A, C and E) as $Pe_s$ is  increased further, i.e., $Pe_s>20$. 

The two-body force also shows a rather complex alternating behavior, when the colloidal distance is fixed and the rescaled channel width is varied. This behavior is shown in Fig. \ref{fig:fig6}, where we have fixed $Pe_s = 50$ and $\tilde \Delta = 0.6$ and plot $\hat F_2$ as a function of $\tilde H$. The origin of this behavior should be traced back to the flat boundary layers of swimmers formed at the two channel walls and the overlaps they produce with the circular layers (rings) formed around the colloids. These overlaps can be  quite prominent and result in an  intriguing pattern of highly populated swimmer regions (appearing as white/yellow spots, layers and arcs  in the panels shown in Fig \ref{fig:fig7}), reminiscent of wavelike interference patterns. As seen in the bottom-most panel for $\tilde H=50$, there are quite a few flat layers at the walls and around the colloids (up to seven layers may be discernible at each of the channel walls and three rings around each of the colloids). As $\tilde H$ is decreased, the rings around the colloids become further enhanced (more strongly populated by the swimmers), while the flat layers at the wall become more suppressed, especially at distances further away from the central colloids. Due to their complexity, the full understanding of the impact of these overlapping patterns remains to be explored. 

The two-body force profiles shown as functions  of the surface-to-surface distance in Figure \ref{fig:fig8} indicate that the effects of the channel width on the force diminish already for $\tilde H=30$ and 40, where the force values nearly saturate. The confinement effects can be considered as substantial for $\tilde H\gtrsim 20$. Thus, while strengthening the confinement initially increases the forces in a quantitative and monotonic fashion across all distances, $\tilde \Delta$, the effects become of qualitative nature  in very narrow channel, e.g., for $\tilde H=15$.

\subsection{Perpendicular orientation}
\label{sec:Vert}

To illustrate the intriguing aspects of the attractive and repulsive forces mediated by the swimmers, we proceed by considering a perpendicular configuration of colloidal particles confined in the channel. To be consistent with the previous section, we consider both low and moderately high occupied area fraction for the bath particles. 

\subsubsection{Low area fraction $(\phi a^{2} = 0.1)$}
\label{sec:rho0p1Vert} 

In the perpendicular position (see, e.g., Fig. \ref{fig:fig2}), the fixed colloidal inclusions in the bath  behave like a barrier to the swimmers. In Fig.~\ref{fig:fig9}, we show the simulated effective two-body force acting on the inclusions as a function of rescaled channel height, $\tilde H$, for different values of the P\'eclet number. The surface-to-surface distance of the colloids is kept fixed as $\tilde \Delta = 2$ (being equal to the diameter of an active particle). Because of the symmetric positional arrangement of the inclusions within the channel, the net (statistically averaged) force acting on the two disks are equal with the $x$ (parallel-to-channel) component averaging out to be zero. Thus, Fig.~\ref{fig:fig9} shows only the $y$ component of the rescaled net force acting on each of the inclusions due to the active particles. As seen in the figure, the force induced by the active particles on the colloidal inclusions is attractive, which is in contrast with the predominantly repulsive force predicted to occur in the bulk system \cite{Harder:JCP2014,Zaeifi:JCP2017}.

\begin{figure}
\centering
\includegraphics[width=1.0\linewidth]{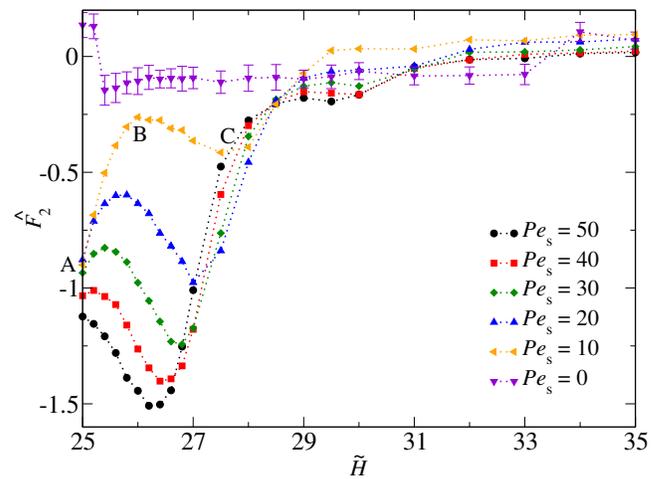}
\caption{
Rescaled effective two-body force, $\hat F_2$, acting on each of the two colloidal inclusions immersed perpendicularly in an active bath is shown as a function of rescaled channel height, $\tilde H$, for different values of the P\'eclet number and at fixed $\tilde \Delta = 2$ and $\phi a^{2} = 0.1$. Symbols are simulation data and lines are guides to the eye. The situations corresponding to the points A, B and C are discussed in the text. The error bars in all cases are typically of the same size as shown for the nonactive case (purple symbols).
}
\label{fig:fig9}
\end{figure}

\begin{figure}
  \includegraphics[width=1.0\linewidth]{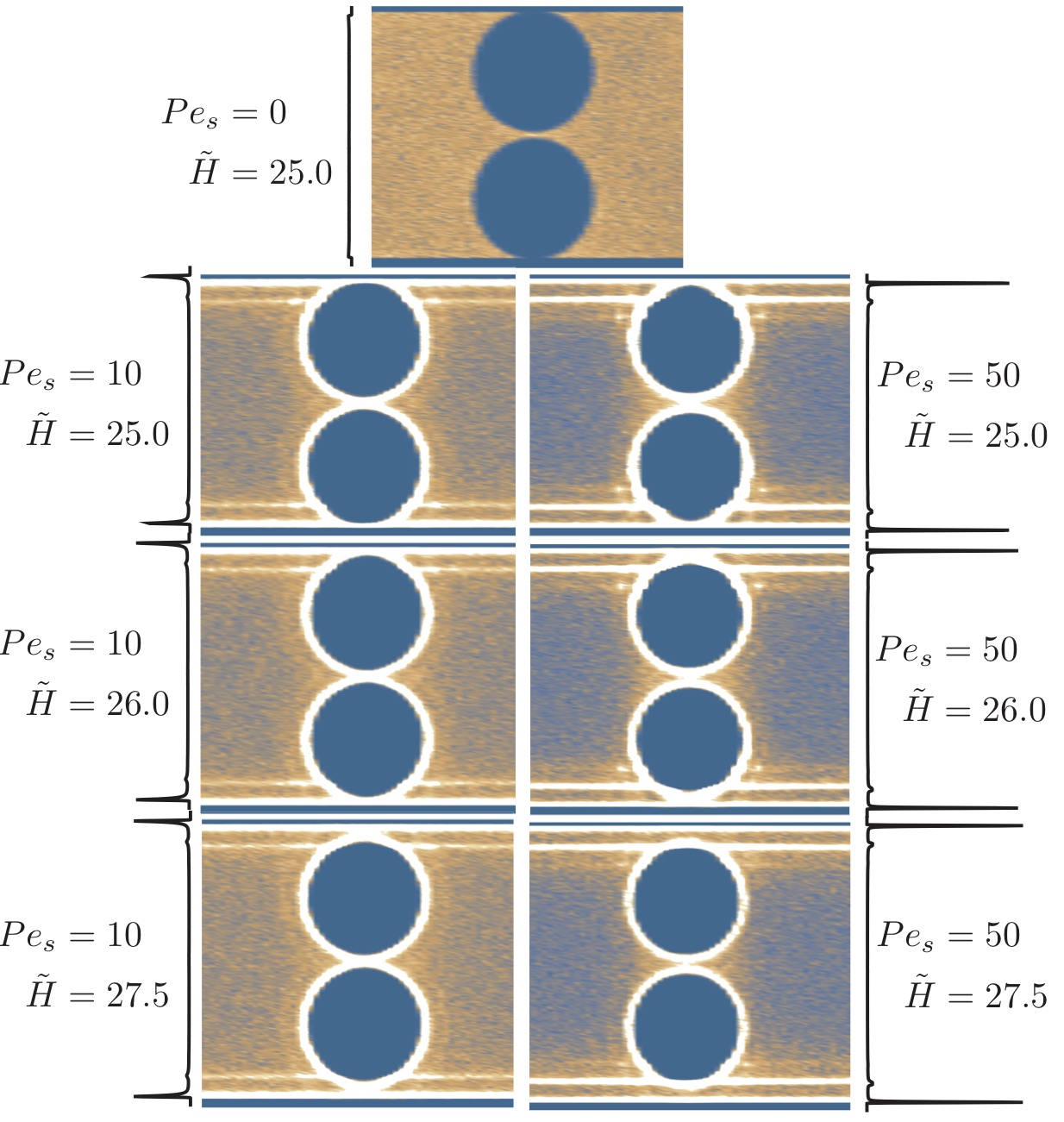}
\caption{Steady-state density map of active swimmers with $Pe_s = 0$ (top) and $Pe_s = 10$ and $50$ (bottom), around two colloidal inclusions located perpendicularly with $\tilde \Delta = 2$, $\tilde H$,  and $\phi a^{2} = 0.1$.
}
\label{fig:fig10}
\end{figure}

The two-body force exhibits a nonmonotonic, oscillating, behavior, with the force magnitude exerted on the colloidal disks showing large variations in narrow channels (e.g., $\tilde H<28)$ at fixed bath activity strength, where it also shows an overall increase with the bath activity strength. These features indicate swimmer structuring around the colloidal disks \cite{Zaeifi:JCP2017}. 
Let us focus on the exemplary case of $Pe_s = 10$ (orange left-triangles in Fig.~\ref{fig:fig9}), in which case the point marked A in the figure is the one giving the  maximum magnitude of the attractive force. This point corresponds to the channel width of $\tilde H=25$ and, since the inter-colloid gap is fixed as $\tilde \Delta=2$, the narrow upper (lower) gap between the top (bottom) colloid of the top (bottom) channel wall is only $3/2$ of a single swimmer radius.  In the cases like this where the surface-to-surface distance between the colloidal inclusions stays constant, the main factor influencing  the effective two-body force is the width of the upper and lower gaps. Any swimmer attempting to go through these narrow passages can squeeze in the first circular swimmer layer that forms around the colloidal inclusions (see Fig.~\ref{fig:fig10}), producing a large attractive force between the colloids. In other words, this force is mediated by the intersection between 1st wall layers and the colloidal surfaces. As the channel gets wider, the induced force on the colloidal inclusion decreases until there is enough space between layers of colloidal inclusions and the wall to fit a complete swimmer particle (point B is the distance that $\tilde H = 26$). After this points, the wall layers moves away from colloidal surfaces and therefore, their overlap gets weaker and as a result, the attractive force mediated from their intersection become weaker. However, the attraction force between colloidal inclusions increases as the height of the channel increases and the particles are trying to enter to the gap between the formed layers around colloids and walls until point C in the figure and starts to decrease as the gap gets wider till the second layers can form for the walls. As expected, the force applied by the intersection between first layers of wall and the colloidal rings has a higher effect comparing to the second wall layers because of their lower swimmer population. This behavior shifts to closer distances as the value of the $Pe_s$ increase which is mainly because a stronger overlap can be possible for bath particles.

Figure 
\ref{fig:fig10} shows the steady state density plot along with the {\it y} density profile corresponding to $Pe_s = 10$ and $Pe_s = 50$  as two examples representing the points $(A, B, C)$ from Fig.~\ref{fig:fig9}. In the case of passive bath particles, there is no layer formation around colloidal inclusions and the wall of the channel and the bath represent a homogeneous medium and the {\it y} profile shows a uniform distribution of the swimmers along the channel width. However, by increasing the activity of bath particles and creating active swimmers, as expected, the density plots show an aggregation of particles around colloidal inclusions and wall of the channel. 
The second and third layers for the walls and the second layer of the colloids have formed around colloidal inclusions and the wall of the channel, respectively. At low P\'eclet number (e.g. $Pe_s = 10)$, the population of swimmers in the area between two walls is high and the second layer can be formed around the colloidal inclusions (the second layer is formed for the walls as well but not its third layer). However, by increasing the P\'eclet number, the second layer around the colloidal inclusions disappears and instead, the third layer forms for the walls. Migration of swimmers from the colloidal layers to the channel layers can be detected from this figure. The \textit{y}-profile shows how the height of the peaks of the wall layers increases as the P\'eclet number increases. In addition, for a given P\'eclet number, as expected, the height of the peaks (population of microswimmers) increases as the channel height increases. 

\subsubsection{High area fraction $(\phi a^{2} = 0.3)$}
\label{sec:rho0p3Vert}

We now consider the same system as above but with a larger area fraction, $\phi a^{2} = 0.3$, occupied by the swimmers. Figure~\ref{fig:fig11} shows the results for the effective two-body force acting on inclusions as a function of channel rescaled height, $\tilde H$  for the case where $Pe_s = 50$. The oscillating force is created between two colloidal inclusions and the attractive part of the force gradually decays as the channel height increases. However, the magnitude of the repulsive part stays almost the same and channel height has less effect on the repulsive part of the force. This is because of the constant surface-to-surface distance that has been chosen for this case. The repulsive force between colloidal inclusions is dominated by the swimmer particles that locate themselves in the area between the two colloidal disks. Since this distance is kept constant the magnitude of the maximum repulsive force stays unchanged. However, the attractive force comes from the particles that is located in the areas between colloidal inclusions and the walls. By increasing this distance the magnitude of maximum attractive force decreases as channel gets wider where no repulsive force is induced from the wall-trapped swimmers. The repulsive force is dominated by the intersection of the second and first rings of one colloid with the surface of the other colloids. Since the number of the rings around colloids stays unchanged as function of channel diameter, therefore, their repulsive interaction stays unchanged. However, the attractive force is dominated by the channel layers-colloid surface intersection, which this decays as channel gets wider (see Fig.~\ref{fig:fig12}).

\begin{figure}
\includegraphics[width=1.0\linewidth]{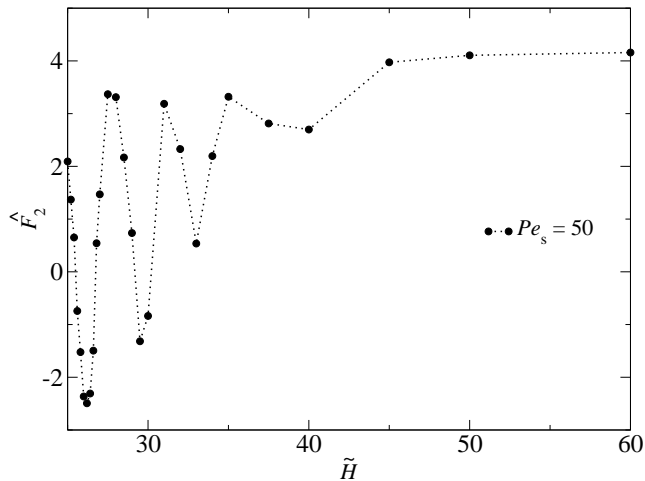}
\caption{Rescaled effective two-body force, $\hat F_2$, acting on each of the two colloidal inclusions immersed perpendicularly in an active bath is shown as a function of channel height, $\tilde H$, for P\'eclet number, $Pe_s = 50$, $\tilde \Delta= 2.0$, and $\phi a^{2} = 0.3$. Note that the plot shows effective force divided by $Pe_s+1$ as defined in the text. Symbols are simulation data and lines are guides to the eye.
} 
\label{fig:fig11}
\end{figure}

Figure \ref{fig:fig12} shows some of the corresponding steady-state density plots from Fig.~\ref{fig:fig11} as examples. As a result of the high concentration of active particles in the bath, even in the case of no colloidal inclusions, there is a high number of layers are formed around the wall of the channel (not shown). By introducing the colloidal particles in a perpendicular orientation, they behave as a barrier and dynamic cluster structures forms around them. Even though at lower channel wall separations the multilayers are formed around colloidal disks, however, by moving two walls away from each other, due to the higher affinity of swimmers to the wall rather than the convex surface of the colloids, number of layers of the colloids decreases and the number of layers for the walls increases. The {\it y}-profiles shows this migration of layers from colloids to the walls as the heights of the peaks increases as channel gets wider.  

\begin{figure}
\includegraphics[width=6.0cm]{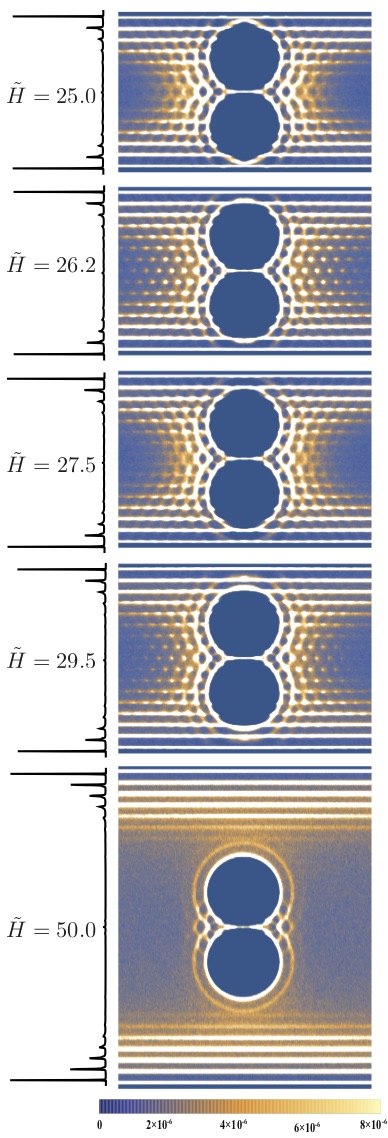}
\caption{Steady-state density map of swimmers with $Pe_s = 50$, around two colloidal inclusions located perpendicularly with $\tilde \Delta = 2$ in channels of different heights, $\tilde H$, as indicated on the graph, at fixed $\phi a^{2} = 0.3$. 
} 
\label{fig:fig12}
\end{figure}

\section{Conclusion}  
\label{sec:Conclusion}

We have studied interactions between colloidal inclusions in a confined two-dimensional geometry containing a bath of self-propelled active particles. The effect of confinement on the effective interaction induced by swimming bath particles on colloidal particles as a function of the orientation of colloidal particles, the magnitude of confinement and the strength of propelling force for swimmers have been investigated. Our results show that the confinement has a strong effect on the interactions between colloidal particles, however, this mainly depends on the colloidal orientation inside the channel. Effect of confinement on the interaction of colloidal disks is more dominant as the self-propulsion increases.

In a narrow channel, and unlike the bulk case \cite{Harder:JCP2014,Zaeifi:JCP2017},  the orientation of colloidal particles plays a critical role in the force applied to them from active bath particles. Since the active bath is homogeneous and isotropic, these forces sum up to zero in the case of a single inclusion. In the presence of a second colloidal inclusion, the spatial isotropy around the first one is broken, giving rise to finite net forces of equal magnitudes (to within the simulation error bars) acting on the inclusions along their center-to-center axis (or the $x$-axis as shown in Fig. \ref{fig:schem_colloids}). In the case of nonactive (nearly-hard) particles, this effective {\em two-body interaction force}, which also represents the potential of mean force between the inclusions \cite{chandler:book1987}, becomes sizable only at sufficiently small separation distances and, as noted previously, originates from the particle depletion effects. The interaction profile in an active bath exhibits alternating behavior, whose features and origin have been investigated in our study. 

Other  interesting problems that can be explored in the present context in the future include the role of  particle-wall and inter-particle hydrodynamic couplings  \cite{BaskaranPNAS2009,StarkPRL2014,wallattraction,ardekani,stark-wall,Hernandez-Ortiz1,elgeti2016} as well as particle chirality and the effects of active noise in the dynamics of the self-propelled particles \cite{Romanczuk:PRL2011,Grosmann:NJPhys2013,Romanczuk:EPJ2015}. 

\section{Acknowledgements}
\label{sec:Acknowledgments}
The computational calculations are provided by the High Performance Computing center of the Institute for Research in Fundamental Sciences (IPM). 
 A.N. acknowledges partial support from the Associateship Scheme of The Abdus Salam International Centre for Theoretical Physics (Trieste, Italy).

\bibliography{ref}

\end{document}